\documentclass[aps,pra,twocolumn,showpacs]{revtex4}
\usepackage{amsmath,graphicx,epsfig}

\newlength{\defbaselineskip}
\newcommand{\setlinespacing}[1]%
          {\setlength{\baselineskip}{#1 \defbaselineskip}}

\begin{document}

\preprint{}

\title{Controlling atomic vapor density in paraffin-coated cells
       using light-induced atomic desorption}
\author{T.~Karaulanov}\email{karaulanov@berkeley.edu}
\author{M.~T.~Graf}
\author{D.~English}
\author{S.~M.~Rochester}
\author{Y.~J.~Rosen}
\author{K.~Tsigutkin}
\author{D.~Budker}\email{budker@berkeley.edu}
\affiliation{Department of Physics, University of California at
Berkeley, Berkeley, California 94720-7300}

\author{E.~B.~Alexandrov}
\author{M.~V.~Balabas}
\affiliation{S. I. Vavilov State Optical Institute, St.
Petersburg, 199034 Russia}

\author{D.~F.~Jackson~Kimball}
\affiliation{Department of Physics, California State
University~--~East Bay, Hayward, California 94542-3084}

\author{F.~A.~Narducci}
\affiliation{EO Sensors Division, US Naval Air Systems Command,
Patuxent River, Maryland 20670}

\author{S.~Pustelny}
\affiliation{Centrum Bada\'n Magnetooptycznych, M.~Smoluchowski
Institute of Physics, Jagiellonian University, Reymonta 4, 30-059
Krak\'ow, Poland}

\author{V.~V.~Yashchuk}
\affiliation{Advanced Light Source Division, Lawrence Berkeley
National Laboratory, Berkeley, California 94720}

\date{\today}

\begin{abstract}
Atomic-vapor density change due to light induced atomic desorption
(LIAD) is studied in paraffin-coated rubidium, cesium, sodium and
potassium cells. In the present experiment, low-intensity probe
light is used to obtain an absorption spectrum and measure the vapor
density, while light from an argon-ion laser, array of light emitting diodes, or discharge lamp is
used for desorption. Potassium is found to exhibit significantly
weaker LIAD from paraffin compared to Rb and Cs, and we were unable
to observe LIAD with sodium. A simple LIAD model is applied to
describe the observed vapor-density dynamics, and the role of the
cell's stem is explored through the use of cells with lockable
stems. Stabilization of Cs vapor density above its equilibrium value over 25 minutes is demonstrated. The results of this work could be used to assess the use of LIAD for vapor-density control in magnetometers, clocks, and
gyroscopes utilizing coated cells.
\end{abstract}
\pacs{34.50.Dy, 79.20.La}

\maketitle

\section{Introduction}

In glass vapor cells atomic polarization relaxes rapidly when atoms
collide with cell walls.  This relaxation can be reduced by up to
four orders of magnitude by introducing a paraffin coating
\cite{Robinson,Bouchiat,Alexandrov1,Alexandrov2}. Long-lived atomic
polarization (relaxation times of seconds have been observed)
enables extremely sensitive measurements of magnetic fields
\cite{Claude,Alexandrov1,Alexandrov2,NMOR,Alexandrov3}, enhances
nonlinear optical effects at low light powers (see the review
\cite{NMOEreview} and references therein), and may make possible
precision tests of fundamental symmetries
\cite{UW_edm,Yashchuk,Kimball}. In addition to these applications,
paraffin-coated cells have been used in the study of light
propagation dynamics \cite{SlowLight,FastLight}, for generation of
spin-squeezed states \cite{Bigelow}, quantum memory for light
\cite{Polzik1}, quantum teleportation \cite{Polzik2}, and creation
and study of high-rank polarization moments
\cite{HexaDecapole,AcostaEarthHexadecapole}. There has also been
renewed interest in the application of paraffin-coated cells in
miniaturized atomic clocks, magnetometers, and gyroscopes
\cite{NISTpaper,Bal2006}. In spite of their wide and varied
application and several detailed studies of their spin-relaxation
properties
\cite{BouchiatPhD,Bouchiat,Gibbs,Liberman,BalabasRID,Vanier,AleksandrovK},
there is still much to learn about the paraffin-coated cells.

In this work, we investigate a possibility to affect density changes
in coated cells using light-induced atomic desorption (LIAD) \cite{PDMS_gozzini,PDMS_mariotti}. The effect
was first investigated in paraffin-coated cells in
Refs.~\cite{ourLIAD,RelaxAtomPolar}. Compared to the earlier work,
we extend the spectral range of the desorbing light into the
ultraviolet, and look at LIAD with potassium (K), comparing the
effect to that with Rb and Cs. We also test for LIAD in sodium (Na).
Using cells with lockable stems (uncoated side-arms containing
metallic alkali samples), we elucidate the role of the stem in the
LIAD dynamics.

LIAD is a process in which alkali atoms are desorbed from the walls
of  a (in our case, coated) vapor cell into the volume of the cell
when the cell is exposed to light of sufficiently short wavelength
and sufficient intensity. This is possible because, over time, alkali atoms are adsorbed into the
coating, and can be released by the action of the desorbing light.
The LIAD phenomenon has been observed using a wide range of
surfaces, besides paraffin: sapphire
\cite{AlexandrovLIAD,BonchB1,BonchB2}, silane-coated glass (in
particular, polydimethylsiloxane (PDMS) )
\cite{PDMS_gozzini,PDMS_meucci,PDMS_mariotti,PDMS_xu,PDMS_atutov,PDMSSilvia},
superfluid $^4$He films \cite{Yabuzaki,YabuzakiNew}, quartz
\cite{QuartzLIAD}, porous silica \cite{PorousSilicaLIAD}, and
octadecyltrichlorosilane (OTS) \cite{MoiOTS,Seltzer}. Rubidium LIAD
from a octadecyldimethylmethoxysilane (ODMS) coating  within a
photonic band-gap fiber was also demonstrated \cite{GaetaPBG}
allowing for realization of efficient nonlinear interactions (in
this case, electromagnetically induced transparency) at ultra-low
optical power. Paraffin coatings are particularly useful because of
the aforementioned long spin relaxation times, while LIAD is an
attractive method for rapid control of atomic density.

The change in vapor density produced by LIAD in a paraffin coated
cell depends on many factors including the cell's geometry, the
wavelength of the desorbing light, the alkali vapor used and the
cell's history, see, for example, Ref.~\cite{ourLIAD}. Important
geometrical factors are the volume, surface area, and stem-opening
area. The latter is of importance because of the so-called
``reservoir" effect \cite{Bouchiat}. The stem of the cell acts
either as a source or a sink for atoms in the volume of the cell
depending on whether the vapor density in the volume is lower or
higher than that in the stem. Thus the ratio of dimensions,
particularly that of the area of the stem opening to the surface
area of the cell, dictates, to a large degree, the time scale for
which the excess density created by LIAD will persist before atoms
retreat to the stem. The reservoir effect is most pronounced in
small cells since they have the smallest internal surface area but
often have stem openings that are comparable to those of larger
cells, so that the stem-opening to surface-area ratio is particulary
unfavorable.

Some applications such as compact, sensitive atomic magnetometers,
clocks, and gyroscopes require high vapor densities. This is because
small cells require higher densities to achieve reasonably sized
signals. In these cases, heating is of limited use with paraffin
coated cells since the temperature cannot exceed $60-80^\circ$C
without melting the coating.  Is LIAD, then, a possible solution for
achieving high density? If so, what is to be done about the
reservoir effect? As a result of the present investigation, the
answer is affirmative, particularly with UV light for
high-efficiency LIAD, and using lockable-stem cells to mitigate the
reservoir effect.

The structure of the paper is as follows: parameters describing the
LIAD dynamics along with a simple model of LIAD processes are
summarized in Sec.~\ref{Sec:LIADyn}; the experimental apparatus are
described in Sec.~\ref{Sec:Setup}; the effect is studied for the
visible desorbing light in Rb and Cs in Sec.~\ref{Sec:CsRbVIS} and
in K and Na in Sec.~\ref{Sec:KNaVIS}; the case when UV irradiation
is used in K and Rb cells is explored in Sec.~\ref{Sec:KRbUV}; tests
with different paraffin coatings are presented in
Sec.~\ref{Sec:Paraffins}. In Sec.~\ref{Sec:ControlCs} demonstration of the control over the Cs density using LIAD is performed. Finally, discussion and conclusion are
presented in Sec.~\ref{Sec:Discuss}.

\section{\label{Sec:LIADyn}LIAD dynamics and its characterization}

\begin{figure}
\includegraphics*[width=3.3in]{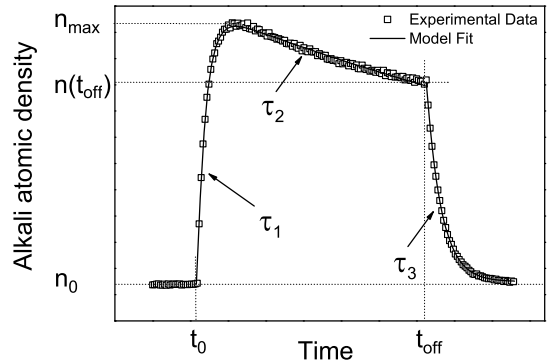}
\caption{Typical LIAD signal and a model fit. The
times when the desorption light is turned on and off are denoted as
$t_{0}$ and $t_{\textrm{off}}$. The initial $n_{0}$, maximum $n_{\textrm{max}}$ and
$n(t_{\textrm{off}})$ densities are indicated. The three characteristic
timescale constants are $\tau_{1}$, $\tau_{2}$ and $\tau_{3}$.}
\label{TeorFit}
\end{figure}

Typical LIAD temporal behavior is illustrated in Fig.~\ref{TeorFit}.
We see three distinct regions: a rapid rise of alkali-vapor density
in the volume of the cell after the desorbing light is turned on at
time $t_0$, a slow subsequent fall-off and a rapid decay of the
vapor density once the light is turned off at time
$t_{\textrm{off}}$.

In Ref.~\cite{ourLIAD}, a model describing most salient features in
the LIAD dynamics was introduced. It is based on the rate equations
and establishment of equilibrium for the atomic vapor density
between the cell volume and the stem before, during and after the
desorbing light action on the paraffin coated alkali cell. A number
of phenomenological parameters are introduced in the model: vapor
density in the stem of the cell, flux of atoms adsorbed into the
coating, exchange rate between the stem and the volume of the cell,
the number of ''free'' atoms in the coating, the rate of
irreversible loss of atoms to the glass or impurities,
light-independent flux of atoms from the surface into the cell
volume, induced desorption rate, etc.

While this model is successful in describing LIAD dynamics in
detail, its significant drawback is the large number of parameters.
Since the focus of the present research is practical applications of
LIAD, we have adopted a greatly simplified version of the model in
which only the most prominent features of LIAD dynamics are
described but the number of fitting parameters is minimized. The
effects that are not accounted for include, for example,
``undershooting'' of the vapor density below its original level
after the desorbing light is turned off. These effects are not
prominent under the experimental conditions employed here.

We define a set of phenomenological parameters that describe the
LIAD dynamics. The maximum change of the atomic density in the
volume of the vapor cell is $\Delta n=n_{\textrm{max}}-n_{0}$, where
$n_{\textrm{max}}$ is the maximum value of the density reached after
the cell is illuminated by desorbing light, and $n_{0}$ is the
initial density prior to illumination. The maximum LIAD yield is
defined as relative change of the density $\eta=\Delta n /n_{0}$.
There are three different time scales for the LIAD dynamics. The
first, shorter time scale with a time constant $\tau_{1}$
characterizes the exponential growth of the alkali-vapor density in
the cell volume just after the desorbing light is switched on. The
second, longer time scale with time constant $\tau_{2}$
characterizes the decrease in density while the desorbing light is
still on. A third time constant $\tau_{3}$ is introduced to account
for density relaxation back to its equilibrium value in the absence
of the desorbing light. Note that in Ref.~\cite{ourLIAD} it was
found that the density drops slightly below the initial value $n_0$
after the light is turned off and eventually recovers on a much
longer time scale. We ignore this effect here. In
Ref.~\cite{ourLIAD} using a more detailed model, it is shown that
the times $\tau_{1}$ and $\tau_{2}$ both depend on the desorbing
light intensity. The loss of atoms from the volume of the cell to
the stem also plays an important role in establishing the time
scales $\tau_1$ and $\tau_2$ \cite{ourLIAD}. In the present work,
using lockable-stem cells, we directly verify this.

We use a simple model to describe the LIAD dynamics, in which the
time evolution of the atomic density in the volume of the cell
$n(t)$ is described by the following equation:

\begin{displaymath}
n(t)=
\begin{cases}
n_{0}+ \textrm{N} \bigl( 1-e^{\frac{-(t-t_{0})}{\tau_{1}}} \bigr)
\: e^{\frac{-(t-t_{0})}{\tau_{2}}}&
\text{, $t_{0} \leq t \leq t_{\textrm{off}}$}\\
n_{0}+ \bigl( n( t_{\textrm{off}})-n_{0} \bigr) \:
e^{\frac{-(t-t_{\textrm{off}})} {\tau_{3}}}&\text{,
$t>t_{\textrm{off}}$}.
\end{cases}
\end{displaymath}

where the relation between $\textrm{N}$ and $\Delta n$ is given by

\begin{displaymath}
\textrm{N} = \Delta n \left( 1+\frac{\tau_{1}}{\tau_{2}} \right)
\left(1+\frac{\tau_{2}}{\tau_{1}}
\right)^\frac{\tau_{1}}{\tau_{2}}.
\end{displaymath}

The moments $t_{0}$ and $t_{\textrm{off}}$ correspond to the times
when the desorption light is switched on and off. The time evolution
corresponding to the equation above is fit to the experimental data,
in this way determining $\tau_{1}$, $\tau_{2}$ and $\tau_{3}$.

\section{\label{Sec:Setup}Apparatus and experimental setup}
\begin{figure}
\includegraphics*[bb=10 10 600 510,width=3.3in]{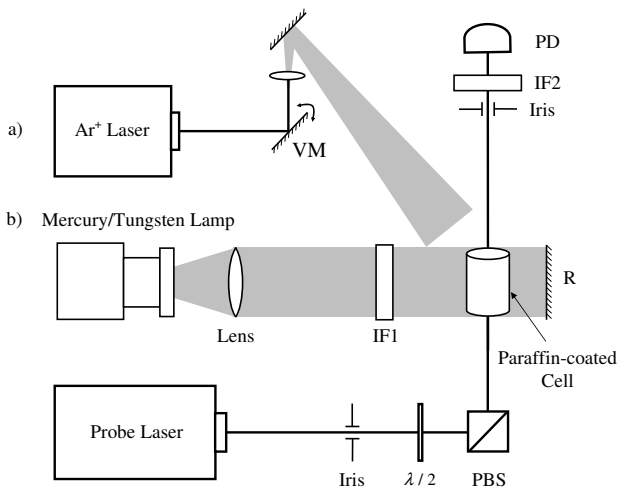}
\caption{Diagram of the experimental setup used in LIAD
measurements. Desorption light in the visible spectrum is generated
by an Ar$^{+}$ laser at 514 nm - case a), while in the UV range
mercury or tungsten discharge lamps are used - case b),
respectively. Half-wave plate ($\lambda /2$) and polarizing beam
splitter (PBS) form a variable attenuator for the probe light.
Interference filters IF1 are used to select the desired desorption
wavelength, while IF2 filters are centered at the alkali atomic
resonance lines and are used to reject the scattered light from the
desorption sources. A vibrating mirror (VM) is used to average the
speckle from the Ar$^{+}$ laser. A reflector (R) increases the
average desorption intensity falling on the surface of the paraffin
cell.} \label{ExpSetup}
\end{figure}

\begin{table}
\caption{\label{tab:table1} Cells used for the LIAD study. For a
given alkali atom the cell stem type, shape, diameter~(D), and
length~(L) are specified.}
\begin{ruledtabular}
\begin{tabular}{cccccccc}
Atom, stem type&Shape&D [mm]&L [mm]\\
\hline
Cs, lockable&cylindrical&$20$&$30$\\
$^{85}$Rb, lockable&cylindrical&$20$&$20$\\
K, lockable&spherical&$50$&$-$\\
Na, non-lockable&cylindrical&$50$&$70$\\
\end{tabular}
\end{ruledtabular}
\end{table}
\begin{figure}
\includegraphics*[width=3.3in]{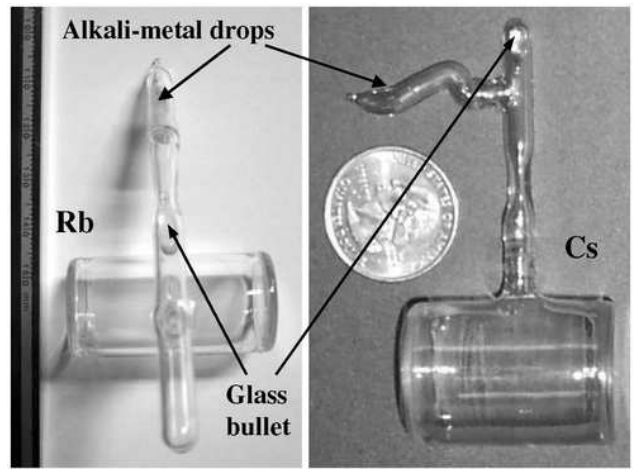}
\caption{Photographs of the Rb and Cs cells with lockable stems.}
\label{RbCsCells}
\end{figure}
The parameters of the paraffin-coated  alkali-vapor cells used in
this work (including the ones shown in Fig.~\ref{RbCsCells}) are
summarized in Table.~\ref{tab:table1}. The stem openings are
circular with a diameter of $\sim 1~$mm. The cells contain a drop of
an alkali metal in their stems. In the case of lockable-stem cells,
a freely moving glass bullet inside the stem is used to open and
close the stem opening by rotating the cell $180^ \circ$ about its
axis and allowing the bullet to slide due to gravity. Special mounts
were designed to allow for consistent locking and unlocking of the
stems. A description of the procedure of coating and filling the
cells can be found in Ref.~\cite{ourLIAD}. Figure~\ref{ExpSetup}
illustrates the general setup of the experiment. For all the cells
there is a probe beam which is resonant with the transition of the
particular alkali atom. For the Cs cell, the beam which is resonant
with the Cs D2 transition originates from a 852-nm extended-cavity
diode laser. For the Rb cell, the beam resonant with the Rb D1
transition is derived from a 795-nm EOSI 2010 extended-cavity diode
laser system. The probe laser for the K~{D1} line is a 770~nm - New
Focus Velocity diode-laser systems. Typical light power of the probe
beams is $5~{\rm \mu W}$ and the diameter is $\approx 3~{\rm mm}$.
In the case when Na is studied a collimated beam originated from a
Na hollow cathode lamp is used. The  probe beam passes through the
cell and falls on a photodiode outfitted with an band-pass
interference filter with a 12-nm FWHM centered near the transition
wavelengths which prevents detection of scattered light from the
light sources used for desorption. Scanning the probe laser (in all
the cases except for Na), we record absorption profiles that are
subsequently analyzed to extract the alkali-vapor density in the
cells. The Ar$^+$ laser (operating at the $514~{\rm nm}$ line), is
employed for desorption in the visible range (see
Fig.~\ref{ExpSetup}a). The desorbing laser beam is reflected from a
vibrating mirror which serves to average the interference pattern in
the laser light (speckle), and expanded using lenses in order to
illuminate the entire cell. The total intensity of the desorption
light (including the retroreflection) incident on the cell ranged
from $1.1~{\rm mW/cm^2}$ to $110~{\rm mW/cm^2}$. The LIAD
experiments in the UV range were performed using 365-nm light
produced by a mercury lamp (Blak-Ray B100 AP) shown on
Fig.~\ref{ExpSetup}b. A McPherson 661A-3 high-pressure tungsten-lamp
system (Fig.~\ref{ExpSetup}b) was used for UV excitation at 313~nm.
Interference filters centered at 365-nm {(FWHM=60~nm)} and 313-nm
{(FWHM=25~nm)} were used to select the desired wavelength.
\section{LIAD in R\MakeLowercase{b} and C\MakeLowercase{s} cells with
Lockable Stems and visible light} \label{Sec:CsRbVIS}

In order to circumvent the reservoir effect, cells with lockable
stems were used. The stems in these cells can be opened or closed
externally, without opening the cell. In the open-stem
configuration, the cell approximates a typical paraffin-coated cell
with a stem, while in the closed configuration it approximates a
cell with no stem. We have experimented with Rb and Cs.

\begin{figure}
\includegraphics*[width=3.3 in]{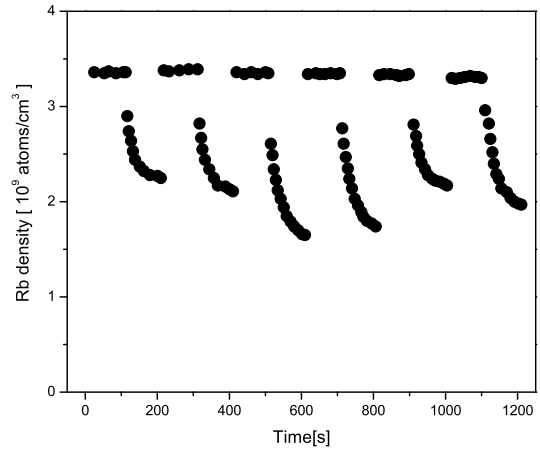}
\caption{Change of the Rb vapor density in the volume of the
lockable-stem cell with the stem alternately open and closed. Higher
densities correspond to the open configuration. The difference in
the closed-stem equilibrium densities is probably due to a variation
of how tightly the stem is closed with different orientation of the
glass bullet.}\label{openclosed}
\end{figure}

\begin{figure}
\includegraphics*[width=3.3 in]{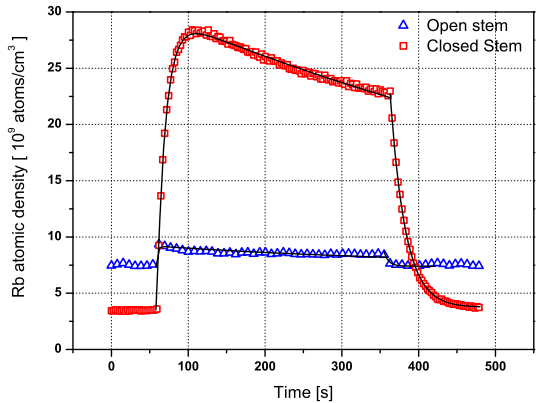}
\caption{(Color online) Density change with LIAD in the Rb
paraffin-coated lockable-stem cell. Two cases are shown: open stem
and closed stem configurations. The solid lines represent the fits
to the data. The applied desorbing light is with intensity of
$50~$mW/cm$^2$ in both cases. The desorbing light was turned on
(around 60 s) and then turned off (around 360~s).} \label{RbLiad}
\end{figure}

For these cells, we sought first to gather evidence of the
reservoir-effect dependence on the area of the stem opening. To do
this, we alternately opened and closed the stem of the Rb cell
(without using LIAD) and recorded the corresponding vapor densities
(Fig.~\ref{openclosed}). It is immediately apparent from the graph
that there is a time dependent density drop once the stem is closed
which we attribute to adsorption of alkali atoms into the coating.
After a number of repetitions, the density with the stem open
remains constant at about $3.4\times 10^9$~atoms/cm$^3$. The density
with the stem closed, however, appears to have different modes (one
at $\sim 2.2\cdot 10^9$~atoms/cm$^3$, another at $\sim 1.8\times
10^9$~atoms/cm$^3$ and the third at $\sim 2\times
10^9$~atoms/cm$^3$). These modes are most likely the result of
different (but relatively stable) configurations of the glass bullet
blocking the part of the stem containing the Rb metal. Similar
behavior is observed with the other lockable-stem cells used.
\begin{figure}
\includegraphics*[width=3.3 in]{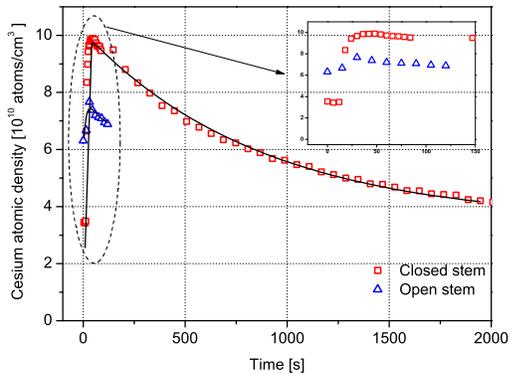}
\caption{(Color online) Change of the atomic density with LIAD
effect in two configuration: open and closed stem in a Cs
paraffin-coated locking-stem cell. The solid line represent a fit to
the data. The applied desorbing light intensity was $28~$mW/cm$^2$
in both cases. The desorbing light was turned on (around t=15~s) and
is never turned off for these measurements.} \label{CsLiad}
\end{figure}
Next, we tested LIAD in Cs and Rb in both open and closed stem
configurations at desorbing laser intensity of $28~$mW/cm$^2$ for Cs
and $50~$mW/cm$^2$ for Rb. The LIAD dynamics in the case of Cs is
presented in Fig.~\ref{CsLiad} and in the case of Rb in
Fig.~\ref{RbLiad}. Model fits are also presented on the figures and
estimated parameters are summarized in Table.~\ref{tab:table2}. As
already commented in Ref.~\cite{ourLIAD}, the stem of the cell plays
a crucial role in the LIAD dynamics and characterization. In both
cells, the closed-stem configuration shows considerably higher
yield of LIAD  for controlling the vapor density in
comparison with the open configuration. The ratio of the yield
$\eta$ for these two configurations is 9 for the Cs cell and 31 for
the Rb cell. Moreover, an important feature is that the rate of
relaxation of the atomic density to its initial value (that before
the desorbing light is turned on) is greatly reduced when the stem
is closed. Both alkalis show $\tau_{2}$ values of the order of 1000
seconds with closed stems, while this time is reduced approximately
by an order of magnitude with open stems.
\begin{table}
\caption{\label{tab:table2}LIAD characterization parameters
extracted from the model fit to the LIAD experimental data. The
maximum LIAD yield $\eta$, the $\tau_{1}$, $\tau_{2}$ and $\tau_{3}$
relaxation times of the atomic density are presented for the Rb and
Cs cells in open (Rb~-~o, Cs~-~o) and closed (Rb~-~c, Cs~-~c) stem
configurations. The desorption light used is characterized by
wavelength~($\lambda$) and intensity~($I$).}
\begin{ruledtabular}
\begin{tabular}{cccccccc}
 Cell&$\lambda$&$I$&$\eta$&$\tau_{1}$&$\tau_{2}$&$\tau_{3}$\\
&(nm)&(mW/cm$^{2}$)&$$&(s)&(s)&(s)\\
\hline
Cs~-~o&$514$&$28$&$0.22$& $18(3)$& $62(6)$&$-$\\
Cs~-~c&$514$&$28$&$1.9$& $7.1(4)$& $869(22)$&$-$\\
Rb~-~o&$514$&$50$&$0.25$& $2.0(3)$& $370(9)$&$6(1)$\\
Rb~-~c&$514$&$50$&$7.7$& $11.8(1)$& $918(9)$&$19.7(1)$\\
Rb~-~o&$365$&$5$&$0.42$& $1.9(3)$& $331(6)$&$1.6(1)$\\
Rb~-~c&$365$&$5$&$7.5$& $13.3(2)$& $1185(29)$&$17.4(3)$\\
Rb~-~c&$313$&$0.25$&$2.4$& $27.9(2)$& $1302(14)$&$32.5(2)$\\
\end{tabular}
\end{ruledtabular}
\end{table}
\section{LIAD in K and N\MakeLowercase{a} cells with visible light}
 \label{Sec:KNaVIS}

An attempt was made to detect LIAD in K. With $30~$mW/cm$^2$ of
514-nm desorption light we were not able to register a change in
potassium density. The upper limit for the weak LIAD effect that may
exist in this case is $\Delta n$$\sim5\times 10^7\ $~atoms/cm$^3$.
Desorbing light of $200~$mW/cm$^2$ caused an increase of the density
of K by $\sim0.2\times 10^9\ $~atoms/cm$^{3}$. A certain amount of
this change of the density may be attributable to heating of the
cell. The LIAD experiment in a Na cell with a normal stem did not
show any observable change of the atomic density. Both alkalis have
low saturated vapor pressure at room temperature in comparison with
Rb and Cs. This may affect the process of ``ripening" of the coating
\cite{ourLIAD, Balabas_Przh}. Perhaps in these cases, the number of free atoms in
the coating ready to be desorbed is small and eventually they are
trapped in sites with deeper interaction potentials. If this is the
case, it explains why we do not observe LIAD from Na and K in the
visible region -- Note, however, that we do observe K LIAD with UV
light (Sec.~\ref{Sec:KRbUV}). Recently, we became aware of a study
\cite{SilviaNaParafin} of LIAD in a Na paraffin coated cell. In this
work, a small change of the sodium density of 2.9$\times
10^8$~atoms/cm$^3$ was observed with 3.5~W/cm$^2$ of desorbing-light
intensity at 514~nm. Note that this intensity is higher by a factor of a hundred than the maximum intensity used (where the data is reliable) in our work. These results are consistent with our observations.

\section{K and R\MakeLowercase{b} LIAD in the Ultraviolet Range}
 \label{Sec:KRbUV}

Earlier work on LIAD from paraffin \cite{ourLIAD} showed that the
efficiency of LIAD increases towards shorter wavelengths of
desorbing light. This important fact may be a clue to better
understanding the processes involved in LIAD. To our knowledge, LIAD
has not been studied in detail for excitation with light of
wavelength shorter than 400 nm. In Ref.~\cite{Moi_UV}, the
desorption dynamics in a PDMS coated Rb cell was explored under the
action of a broad-spectrum light from a mercury-discharge lamp.
However, wavelength selectivity in the UV spectral range was not
specified. Recently, UV light at 395 nm and 253 nm was used for
LIAD-assisted loading of magneto-optical traps \cite{UV_MOT_Trap}
for Rb and K atoms. The authors of this work used uncoated surfaces
- quartz cell walls and the walls of a stainless-steel chamber. Here
we investigate LIAD in paraffin-coated rubidium and potassium cells
with desorbing light at 365~nm and 313~nm. We separately discuss the
study of K and Rb because no strong evidence for existence of LIAD
was found for K with 514-nm excitation with desorbing-light
intensity up to 200~mW/cm$^{2}$. The short-wavelength limit (313~nm)
for the desorption light is due to the absorption of molyglass, the
material of the cells.

\subsection{Case of Rb}

\begin{figure}
\includegraphics*[width=3.3 in]{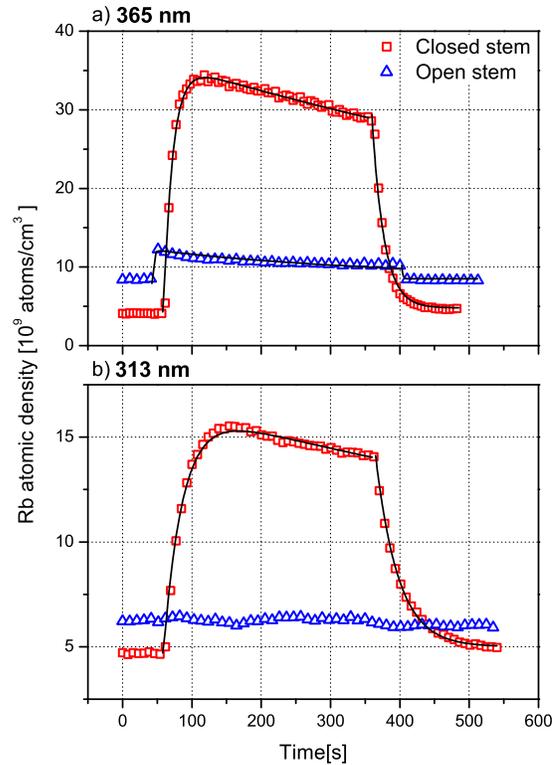}
\caption{(Color online) Atomic-density dynamics in a paraffin coated
$^{85}$Rb cell with lockable stem in case of a) 365~nm desorbing
light with intensity of 5~mW/cm$^{2}$ and b) 313~nm at
0.25~mW/cm$^{2}$. The abrupt kinks in the plots correspond to the
opening and closing of the desorbing light. Note the differences in
the initial densities.} \label{UVRbLIAD}
\end{figure}

The Mercury lamp was first used as the desorption-light source with
the filter centered at 365~nm. We were able to apply up to
$5~$mW/cm$^{2}$ to the Rb cell. The results with the closed and open
stem are presented in Fig.~\ref{UVRbLIAD}a. The parameters from the
model fit are presented in Table.~\ref{tab:table2}. The maximum
observed change in the density in the closed-stem case is from $
4\times 10^9$ to $ 34\times 10^9\ $ atoms/cm$^{3}$. For comparison,
to obtain a similar more than eight-fold increase of atomic density
($\eta =7.5$) using light at 514~nm with the same cell, an order of
magnitude higher intensity of $\simeq 55 $~mW/cm$^{2}$ is needed.
Both $\tau_{1}$ and $\tau_{2}$ times are comparable to the case of
desorption by visible light at 514~nm and intensity of
50~mW/cm$^{2}$. In the open-stem case, the maximum density reached
is less than twice the initial density ($\eta =0.42$). In
Fig.~\ref{UVRbLIAD}b, the excitation with 313-nm light (FWHM=25~nm)
with intensity of only $0.25~$mW/cm$^{2}$ is presented. In the case
when the stem is open, the LIAD effect is undetectable. However,
when the stem is closed, strong increase of atomic density ($\eta
=2.4$) is achieved. The dynamics in both cases of UV excitation is
adequately described with the model used for the visible range.
\subsection{Case of K}

\begin{figure}
\includegraphics*[width=3.3 in]{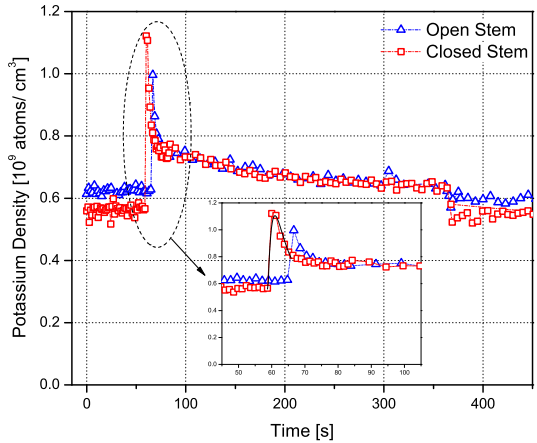}
\caption{(Color online) Atomic density dynamics in the
paraffin-coated K cell with lockable stem in case of 365-nm
desorbing light with intensity of 8~mW/cm$^{2}$. The horizontal
offset of the peaks is due to the difference in the time when the
desorbing light was applied.} \label{UVKLIAD}
\end{figure}

The Mercury lamp was used as the source of desorbing light with the
filter centered at 365 nm and intensity at the cell of $ \sim 8\
$mW/cm$^{2}$. The experimental results are presented in
Fig.~\ref{UVKLIAD}. In contrast with the case when 514-nm desorbing
light was used, there is an observable LIAD with UV light. This once
again illustrates the fact that the efficiency of LIAD increases
towards UV. In the case of K, in contrast to all other data, our
model does not describe the entire temporal profile of the vapor
density. However, fitting the fast rising and falling slopes only
(closed stem configuration), we extract the maximum yield
$\eta$=2.3, $\tau_{1}$=$1.7(4)$~s and $\tau_{2}$=$4$~s. Note a
drastic difference in the LIAD dynamics from what is observed with
Rb: the times $\tau_1$ and $\tau_2$ are both much shorter, and there
is no significant difference between the open- and closed-stem
cases.
\section{LIAD Performance of different types of Paraffin coatings}
 \label{Sec:Paraffins}

We tested non-lockable-stem cells containing two different types of
paraffin wax. This experiment was performed to determine to which
extent LIAD parameters depend on the specific type of paraffin used
for coating. There are four 3.5-cm diameter spherical cells in this
experiment which we will label cells 1--4, respectively. All these
cells were made using the method described in Ref.~\cite{ourLIAD}
and contain Rb with natural isotopic abundance. Cells 1 and 2 were
coated with the same paraffin as used in the lockable-stem cells and
the cells used in Ref.~\cite{ourLIAD}. Cells 3 and 4 were coated
with a different paraffin (Luxco wax F/R130). The cells were exposed
to 514-nm light with an intensity of $74\ $mW/cm$^2$. On the first
trial only Cell 1 showed any response to the desorbing light. After
baking the cells overnight at $70^\circ\ $C all the cells showed
significant improvement of their LIAD properties. The results are
summarized in Table.~\ref{tab:table3}. The results indicate that
both paraffins are similar in their LIAD properties. However, the
process of ``ripening" plays an essential role in the desorption
properties of the coating. This leads to a conclusion that cell
preparation and history can significantly change the observed LIAD
properties.
\begin{table}
\caption{\label{tab:table3} LIAD in Rb cells with different paraffin
coatings. The cell number; paraffin type; the initial density of
atoms $n_{0}$; the maximum change of the atomic density $\delta n$;
and the maximum LIAD yield $\eta$ are listed. Desorbing light at
514~nm with intensity of $74~$mW/cm$^2$ was used.}
\begin{ruledtabular}
\begin{tabular}{cccccccc}
Cell number&Paraffin type&$n_{0}$&$\delta n$&$\eta$\\
&$$&(10$^{9}$cm$^{-3}$)&(10$^{9}$cm$^{-3}$)&$$\\
\hline
1&Ref.\cite{ourLIAD}&$2.3$&$1.2$&$0.5$\\
2&Ref.\cite{ourLIAD}&$2$&$1$&$0.5$\\
3&Luxco wax F/R130&$1.8$&$1$&$0.6$\\
4&Luxco wax F/R130&$1.8$&$0.5$&$0.3$\\
\end{tabular}
\end{ruledtabular}
\end{table}

\section{Control of atomic density using LIAD}
 \label{Sec:ControlCs}

\begin{figure}
\includegraphics*[width=3.3 in]{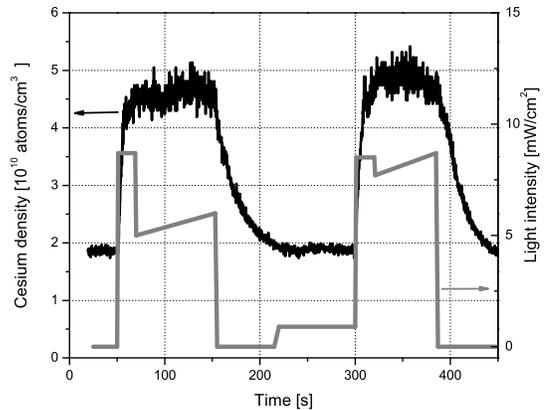}
\caption{Demonstration of switching of Cs atomic density using LIAD
with a lockable-stem cell (stem closed). The intensity of the
desorbing 405-nm light (FWHM=15 nm) is shown on the lower trace.}
\label{LIADPulseStab}
\end{figure}

\begin{figure}
\includegraphics*[width=3.3 in]{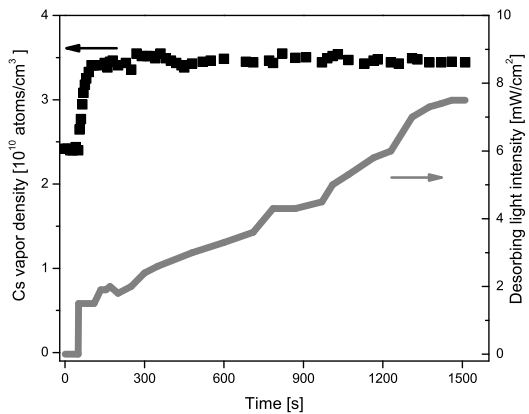}
\caption{Demonstration of Cs atomic density stabilization using LIAD
with lockable-stem cell (stem closed). The intensity of the
desorbing 405-nm light (FWHM=15 nm) is shown on the lower trace.}
\label{LIADStab}
\end{figure}

In order to demonstrate the use of LIAD for controlling atomic
density, we performed two experiments with the lockable-stem Cs
cell. The source of the desorbing light in both cases was an array
of four light-emitting diodes (LEDs) with central wavelength of 405
nm (FWHM = 15 nm). The LED-array light intensity is modulated via
changing the current. In Fig.~\ref{LIADPulseStab} a two-pulse
desorption sequence is demonstrated. In this case, the initial light
intensity for both pulses is the maximum available $\approx 9\
$mW/cm$^2$ used to minimize the density ramp-up time $\tau_{1}$. In
this way, relatively fast switching time of $\approx 2\ $s for the
Cs atomic density change of $\sim 1\times 10^{10}\ $ atoms/cm$^{3}$ is
realized. Unfortunately, there is no analogous control over the
density ramp-down time, which is correspondingly lower, $\approx$ 15
s. Stabilization of the atomic density in the cell (stem closed)
over a period of more then 20 minutes with an instability less than
$1.5\times 10^9\ $ atoms/cm$^{3}$ is shown in Fig.~\ref{LIADStab}.
The lower trace on the plot shows the desorbing-light intensity.
Note that the cell temperature variation during the entire cycle
when the vapor density is increased with the help of LIAD by a
factor of 1.5 is less than $0.5^\circ$C. Varying the cell temperature by $\approx
3^\circ$C would be required to achieve this vapor-density change
thermally.

\section{Discussion and Conclusion} \label{Sec:Discuss}
\begin{figure}
\includegraphics*[width=3.3 in]{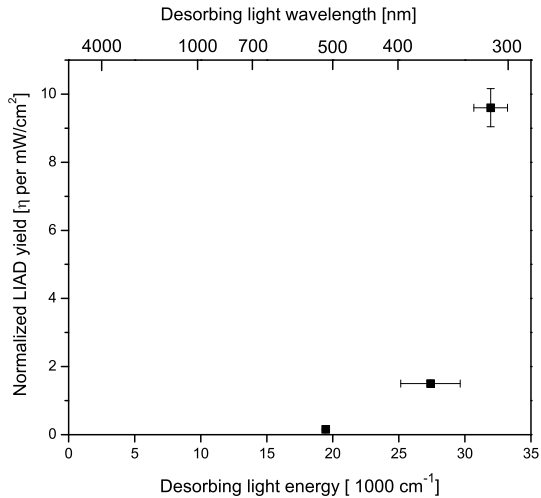}
\caption{Normalized maximum LIAD yield in Rb lockable stem cell
(stem closed) as a function of desorbing light wavelength - 514~nm
(Ar laser), 365~nm (filter FWHM=60~nm) and 313~nm (filter
FWHM=25~nm).} \label{LIADeff}
\end{figure}
We have explored the LIAD effect in paraffin-coated Rb, Cs, K, and
Na cells extending the range of the desorption-light spectrum from
visible to UV. We used lockable-stem cells in order to reduce, to a
large degree, the relaxation of the alkali-atom density due to the
stem. The main advantage of using the lockable-stem cells (stem
closed) is the possibility of obtaining larger densities, for
example, a factor of five gain in comparison with an open stem for
Rb in our experiment. More importantly, one obtains a much longer
density relaxation time ($\tau_{2}$).

In the case of Rb and Cs, atoms with higher saturated vapor pressure
at room temperature, irradiation with a few tens of mW/cm$^2$ of
visible (514-nm) light causes significant LIAD effect. With Na and K
cells and visible desorption light, no LIAD was observed.

Two different paraffins used with Rb do not show significant
difference in the LIAD performance of the cells.

Experiments with UV light (365~nm and 313~nm) and a Rb lockable-stem
cell show particularly high-yield LIAD. When using UV light at
365~nm, an order-of-magnitude gain in the yield of the desorption
process ($\eta$) is realized in comparison with using 514-nm light.
Intensities as low as 0.25~mW/cm$^{2}$ at 313-nm light cause even
stronger LIAD in Rb. Maximum LIAD yield in the Rb lockable-stem cell
(stem closed) normalized to intensity \cite{Normalization} is plotted in
Fig.~\ref{LIADeff} against the wavelength of the desorbing light -
514~nm (Ar laser), 365~nm (filter FWHM=60~nm) and 313~nm (filter
FWHM=25~nm). A strong increase of the yield occurs towards shorter
wavelengths.

In the case of K lockable-stem cell and UV desorbing light (365~nm) we registered much faster relaxation (short $\tau_{2}$ time) of the vapor density compared to the case of Rb and Cs. Note that insufficient ripening is not a likely reason for the difference because K vapor pressure in the cell (open and closed stem configurations) was within 30 percents of saturated K vapor pressure at the same temperature.

We plan to extend the study of LIAD of alkali atoms in
paraffin-coated cells towards the 200-nm range (with cells made of
UV-transmitting glass). This will reveal whether the yield continues
to increase for shorter wavelengths, or whether the presently
observed dependence is the long-wavelength portion of a broad
resonance. Indeed, resonant photodesorption was observed for some
other systems, for example, CO on Ni \cite{Photodesorption}
(resonance centered at 270~nm, FWHM=85~nm).

Combining the increased LIAD efficiency in the UV range with the use
of closed-stem cells opens the door to practical applications, for
example, for stabilizing or modulating the atomic density. Thus we
may replace complicated (for example, non-magnetic) and
high-power-consuming heating systems with a simple LIAD scheme with
desorbing light produced with UV light-emitting diodes. This is
demonstrated in this work with a Cs lockable-stem cell, see Figs.
\ref{LIADStab} and \ref{LIADPulseStab} that show stabilization and
pulsing alkali-vapor density using LIAD. Such fast, high-contrast
control of the vapor density is hard to achieve by heating/cooling
the vapor cell.

LIAD may become a useful technique in the ongoing work aimed at
developing highly miniaturized atomic frequency references
\cite{SmallClocks1,SmallClocks2}, magnetometers \cite{SmallMags},
and gyroscopes. These devices take advantage of miniature atomic
vapor cells with physical dimensions on the order of 1~mm or smaller
\cite{SmallCells, Bal2006}. In order to increase the signal in such
miniaturized cells, an efficient method to elevate the atomic
density is required. As it is shown in Ref.~\cite{RelaxAtomPolar},
compared to changing vapor densities by heating coated cells, LIAD
may offer significant improvement in  terms of spin-relaxation
times.

\begin{acknowledgments}
The authors would like to thank Joseph W. Tringe for providing the
Blak-Ray discharge lamp and Luxco Wax for providing wax samples. This research was supported by the Office
of Naval Research (grants N00014-97-1-0214 and SBIR), the Russian
Foundation for Basic Research (grant 03-02-17509 RFBR), the National
Science Foundation, a Cal-Space Mini-grant, NURI grant
HM1582-08-1-0006, Lawrence Berkeley National Laboratory's Nuclear
Science Division, and the University of California at Berkeley's
Undergraduate Research Apprenticeship Program.
\end{acknowledgments}

\


\begin{thebibliography}{99}

\bibitem{Robinson}
H.\ G.\ Robinson, E.\ S.\ Ensberg, and H.\ G.\ Dehmelt, Bull. Am.
Phys. Soc. {\bf 3}, 9 (1958).

\bibitem{Bouchiat}
M.\ A.\ Bouchiat and J.\ Brossel, Phys. Rev. {\bf 147}, 41 (1966);
M.\ A.\ Bouchiat, Ph. D. Thesis, University of Paris, 1964.

\bibitem{Alexandrov1}
E.\ B.\ Alexandrov and V.\ A.\ Bonch-Bruevich, Opt. Engin.
{\bf{31}}, 711 (1992).

\bibitem{Alexandrov2}
E.\ B.\ Alexandrov, M.\ V.\ Balabas, A.\ S.\ Pasgalev, A.\ K.\ Vershovskii,
 and N.\ N.\ Yakobson, Laser Phys. {\bf{6}}, 244 (1996).

\bibitem{Claude}
J.\ Dupont-Roc, S.\ Haroche and C.\ Cohen-Tannoudji, Phys. Lett.
 {\bf{28A}}, 638 (1969);
C.\ Cohen-Tannoudji, J.\ Dupont-Roc, S.\ Haroche, and F.\ Laloe,
Phys. Rev. Lett. {\bf{22}}, 758 (1969).

\bibitem{NMOR}
D.\ Budker, V.\ Yashchuk, and M.\ Zolotorev, Phys. Rev. Lett.
 {\bf{81}}, 5788 (1998); D.\ Budker, D.\ F.\ Kimball, S.\ M.\ Rochester,
  and V.\ V.\ Yashchuk, Phys. Rev. Lett. {\bf 85}, 2088 (2000);
   D.\ Budker, D.\ F.\ Kimball, S.\ M.\ Rochester, V.\ V.\ Yashchuk, and
M.\ Zolotorev, Phys. Rev. A {\bf 62}, 043403 (2000);
 D.\ Budker, D.\ F.\ Kimball,  V.\ V.\ Yashchuk, and M.\ Zolotorev,
  Phys. Rev. A {\bf 65}, 055403 (2002).

\bibitem{Alexandrov3}
E.\ B.\ Alexandrov, M.\ V.\ Balabas, A.\ K.\ Vershovski, and A.\
S.\ Pazgalev, Technical Physics {\bf{49}}(6), 779-783 (2004).

\bibitem{NMOEreview} D.\ Budker, W.\ Gawlik, D.\ F.\ Kimball,
 S.\ M.\ Rochester, V.\ V.\ Yashchuk, and A.\ Weiss,
  Rev. Mod. Phys. {\bf 74}, 1153 (2002).

\bibitem{UW_edm}
E.\ S.\ Ensberg, Phys. Rev. {\bf 153}, 36 (1967).

\bibitem{Yashchuk}
V.\ Yashchuk, D.\ Budker and M.\ Zolotorev, in
 {\it Trapped Charged Particles and Fundamental Physics}
  AIP Conference Proceedings 457, edited by D.H.E. Dubin and D.
Schneider (American Institute of Physics, New York, 1999), pp. 177-181.

\bibitem{Kimball}
D.\ F.\ Kimball, D.\ Budker, D.\ S.\ English, C.-H. Li, A.-T. Nguyen,
 S.\ M.\ Rochester, A.\ Sushkov, V.\ V.\ Yashchuk, and M.\ Zolotorev,
  in {\it Art and Symmetry in Experimental Physics}
   AIP Conference Proceedings 596, edited by D.\ Budker, P.\ H.\
Bucksbaum, and S.\ J.\ Freedman
 (American Institute of Physics, New York, 2001), pp. 84-107.

\bibitem{SlowLight}
D.\ Budker, D.\ F.\ Kimball, S.\ M.\ Rochester, and V.\ V.\ Yashchuk,
 Phys. Rev. Lett. {\bf 83}, 1767 (1999).

\bibitem{FastLight}
L.\ J.\ Wang, A. Kuzmich, and A. Dogariu, Nature {\bf 406}, 277 (2000).

\bibitem{Bigelow}
A.\ Kuzmich, L.\ Mandel, and N.\ P.\ Bigelow, Phys. Rev. Lett.
 {\bf 85}, 1594 (2000).

\bibitem{Polzik1}
B.\ Julsgaard, J.\ Sherson, J.\ Ignacio Cirac, Jaromir Fiurasek, and E.\ S.\
Polzik, Nature {\bf 432}, 482 (2004).

\bibitem{Polzik2}
J.\ F.\ Sherson, H.\ Krauter, R.\ K.\ Olsson, B.\ Julsgaard, K.\ Hammerer,
I.\ Cirac, and E.\ S.\ Polzik, Nature {\bf 443}, 557 (2006).

\bibitem{HexaDecapole}
V.\ V.\ Yashchuk, D.\ Budker, W.\ Gawlik, D.\ F.\ Kimball,
 Yu.\ P.\ Malakyan, and S.\ M.\ Rochester, Phys. Rev. Lett. {\bf 90}, 253001
(2003).

\bibitem{AcostaEarthHexadecapole}
V.\ M.\ Acosta, M.\ Auzinsh, W.\ Gawlik, P.\ Grisins, J.\ M.\ Higbie,
 D.\ F.\ Jackson Kimball, L.\ Krzemien, M.\ P.\ Ledbetter, S.\ Pustelny,
  S.\ M.\ Rochester, V.\ V.\ Yashchuk, and D.\ Budker,
  http://arxiv.org/abs/0709.4283.

\bibitem{NISTpaper}
D.\ Budker, L.\ Hollberg, D.\ F.\ Kimball, J.\ Kitching, S.\ Pustelny,
and V.\ V.\ Yashchuk, Phys. Rev. A {\bf{71}}, 012903 (2005).

\bibitem{Bal2006}
M.\ V.\ Balabas, D.\ Budker, J.\ Kitching, P.\ D\. D.\ Schwindt, and J.\ E.\
Stalnaker, JOSA B \textbf{23}(6), 1001 (2006).

\bibitem{BouchiatPhD}
M.\ A.\ Bouchiat, Ph. D. thesis, L'Universite de Paris (1964).

\bibitem{Gibbs}
H.\ Gibbs, Phys. Rev. {\bf 139}, 1374 (1965).

\bibitem{Liberman}
V.\ Liberman and R.\ J.\ Knize, Phys. Rev. A {\bf 34}, 5115 (1986).

\bibitem{BalabasRID}
M.\ V.\ Balabas, M.\ I.\ Karuzin, and A.\ S.\ Pazgalev, JETP Lett. {\bf
70}, 196 (1999).

\bibitem{Vanier}
J.\ Vanier, J.-F.\ Simard, and J.-S.\ Boulanger,
 Phys. Rev. A {\bf{9}}(3), 1031 (1974).

\bibitem{AleksandrovK}
E.\ B.\ Aleksandrov, M.\ V.\ Balabas, A.\ K.\ Vershovskii,
 A.\ I.\ Okunevich, and N.\ N.\ Yakobson, Optics and Spectroscopy,
{\bf{87}}(3), 329-34 (1999).

\bibitem{PDMS_gozzini}
A.\ Gozzini, F.\ Mango, J.\ H.\ Xu, G.\ Alzetta, F.\ Maccarrone, and
R.\ A.\ Bernheim,
 Nuovo Cimento D {\bf 15}, 709 (1993).

\bibitem{PDMS_mariotti}
E.\ Mariotti, S.\ Atutov, M.\ Meucci, P.\ Bicchi, C.\ Marinelli, and
L.\ Moi,
 Chem. Phys. {\bf 187}, 111 (1994).

\bibitem{ourLIAD}
E.\ B.\ Alexandrov, M.\ V.\ Balabas, D.\ Budker, D.\ English, D.\ F.\ Kimball, C.-H.\ Li,
 and V.\ V.\ Yashchuk, Phys. Rev. A {\bf 66}, 042903 (2002).

\bibitem{RelaxAtomPolar}
M.\ T.\ Graf, D.\ F.\ Kimball, S.\ M.\ Rochester, K.\ Kerner, C.\ Wong, D.\
Budker, E.\ B.\ Alexandrov, M.\ V.\ Balabas, and V.\ V.\ Yashchuk, Phys.
Rev. A {\bf{72}}, 023401 (2005).

\bibitem{AlexandrovLIAD}
I.\ N.\ Abramova, E.\ B.\ Aleksandrov, A.\ M.\ Bonch-Bruevich,
 and V.\ V.\ Khromov, Pis'ma Zh. Eskp. Teor. Fiz.
  {\bf 39}, 172 (1984) [JETP Lett. {\bf 39}, 203 (1984)].

\bibitem{BonchB1}
A.\ M.\ Bonch-Bruevich, Yu.\ N.\ Maksimov, S.\ G.\ Przhibel'ski\u{i},
 and V.\ V.\ Khromov, Zh. Eskp. Teor. Fiz. {\bf 92}, 285 (1987)
  [Sov. Phys. JETP {\bf 65}, 161 (1987)].

\bibitem{BonchB2}
A.\ M.\ Bonch-Bruevich, T.\ A.\ Vartanyan, Yu.\ N.\ Maksimov,
 S.\ G.\ Przhibel'ski\u{i}, and V.\ V.\ Khromov, Zh. Eskp. Teor. Fiz.
  {\bf 97}, 1761 (1990) [Sov. Phys. JETP {\bf 70}, 993 (1990)].

\bibitem{PDMS_meucci}
M.\ Meucci, E.\ Mariotti, P.\ Bicchi, C.\ Marinelli, and L.\ Moi,
 Europhys. Lett. {\bf 25}, 639 (1994).

\bibitem{PDMS_xu}
J.\ H.\ Xu, A.\ Gozzini, F.\ Mango, G.\ Alzetta, and R.\ A.\ Bernheim,
 Phys. Rev. A {\bf 54}, 3146 (1996).

\bibitem{PDMS_atutov}
S.\ N.\ Atutov, V.\ Biancalana, P.\ Bicchi, C.\ Marinelli, E.\
Mariotti,
 M.\ Meucci, A.\ Nagel, K.\ A.\ Nasyrov, S.\ Rachini, and L.\ Moi,
  Phys. Rev. A {\bf 60}, 4693 (1999).

\bibitem{PDMSSilvia}
S.\ Gozzini and A.\ Lucchesini,
Eur. Phys. J. D {\bf{28}} 157-162, (2004).

\bibitem{Yabuzaki}
A.\ Hatakeyama, K.\ Oe, K.\ Ota, S.\ Hara, J.\ Arai, T.\ Yabuzaki, and
A.\ R.\ Young, Phys. Rev. Lett. {\bf{84}}, 1407 (2000).

\bibitem{YabuzakiNew}
A.\ Hatakeyama, K.\ Enomoto, N.\ Sugimoto, and T.\ Yabuzaki,
 \pra {\bf{65}}, 022904 (2002).

\bibitem{QuartzLIAD}
J.\ Brewer and H.\ G.\ Rubahn, Chem. Phys. {\bf{303}}(1-3), 1-6 (2004).

\bibitem{PorousSilicaLIAD}
A.\ Burchianti, C.\ Marinelli, A.\ Bogi, J.\ Brewer, K.\ Rubahn, H.\ G.\ Rubahn,
 F.\ Della Valle, E.\ Mariotti, V.\ Biancalana, S.\ Veronesi, and L.\ Moi,
  Europhysics Letters {\bf{67}}(6), 983-989 (2004).

\bibitem{MoiOTS}
A.\ Cappello, C.\ de Mauro, A.\ Bogi, A.\ Burchianti, S.\ Di Renzone, A.\
Khanbekyan, C.\ Marinelli, E.\ Mariotti, L.\ Tomassetti, and L.\ Moi,
The Journal of Chem. Physics {\bf{127}}, 044706 (2007).

\bibitem{Seltzer}
S.\ J.\ Seltzer and M.\ V.\ Romalis, Private Communication.

\bibitem{GaetaPBG}
S.\ Ghosh, A.\ R.\ Bhagwat, C.\ K.\ Renshaw, S.\ Goh, A.\ L.\ Gaeta,
and B.\ J.\ Kirby, Phys. Rev. Lett. {\bf{97}}, 023603 (2006).

\bibitem{Balabas_Przh}
M.\ V.\ Balabas and S.\ G.\ Przhibelskii, Chem. Phys. Reports {\bf{14}}
(6), 882-889 (1995).

\bibitem{SilviaNaParafin}
S.\ Gozzini, A.\ Lucchesini, L.\ Marmugi, and G.\ Postorino, Eur. Phys.
J. D {\bf{47}}, 1-5 (2008).

\bibitem{Moi_UV}
E.\ Mariotti, S.\ Atutov, M.\ Meucci, P.\ Bicchi, C.\ Marinelli
and L.\ Moi, Chemical Physics {\bf{187}}, 111 (1994).

\bibitem{UV_MOT_Trap}
C.\ Klempt, T.\ van Zoest, T.\ Henninger, O.\ Topic, E.\ Rasel, W.\ Ertmer,
and J.\ Arlt, Phys. Rev. A {\bf{73}}, 013410 (2006).

\bibitem{SmallClocks1}
J.\ Kitching, S.\ Knappe, and L.\ Hollberg, Appl. Phys. Lett.
{\bf{81}}, 553 (2002).

\bibitem{SmallClocks2}
Y.\ Y.\ Jau, A.\ B.\ Post, N.\ N.\ Kuzma, A.\ M.\ Braun, M.\ V.\ Romalis,
and W.\ Happer, Phys. Rev. Lett. {\bf{92}}, 110801 (2004).

\bibitem{SmallMags}
P.\ Schwindt, S.\ Knappe, V.\ Shah, L.\ Hollberg, J.\ Kitching, L.\
Liew, and J.\ Moreland, Appl. Phys. Lett. {\bf{85}}, 6409 (2004).

\bibitem{SmallCells}
L.\ Liew, S.\ Knappe, J.\ Moreland, H.\ G.\ Robinson, L.\ Hollberg, and
J.\ Kitching, Appl. Phys. Lett. {\bf{84}}, 2694 (2004).

\bibitem{Normalization}
The normalization for the three wavelengths used suppose linear dependence of the yield on the desorption light intensity. We verified the linearity for the case of 365-nm desorpbing light (in the interval of 1 to 10 mW/cm$^2$). In Ref.~\cite{ourLIAD} a nonlinear behavior of the yield is registered at low desorption intensity in open-stem cells. We did not observe such behavior in rubidium lockable-stem cell (stem closed) at 365-nm desorbing light.

\bibitem{Photodesorption}
R.\ O.\ Adams and E.\ E.\ Donaldson, The Journal of Chem. Physics
{\bf{42}} (2), 770-774 (1965).


\end{thebibliography}
\end{document}